\pdfoutput=1

\documentclass{article}
\usepackage{ijcai20}

\usepackage{times}

\usepackage{soul}
\usepackage{url}
\usepackage[hidelinks]{hyperref}
\usepackage[utf8]{inputenc}
\usepackage[small]{caption}
\usepackage{graphicx}
\usepackage{amsmath}
\usepackage{booktabs}
\usepackage{todonotes}
\usepackage{subcaption}
\usepackage{balance}

\title{Detecting Driver's Distraction using Long-term Recurrent Convolutional Network}

\author{
Chang Wei Tan$^1$\footnote{Contact Author}\and
Mahsa Salehi$^1$ \and
Geoffrey Mackellar$^2$\\
\affiliations
$^1$Faculty of Information Technology, Monash University \\
$^2$Emotiv Research \\
\emails
chang.tan@monash.edu,
mahsa.salehi@monash.edu,
geoff@emotiv.com
}

\graphicspath{{figures/}}

\begin{document}

\maketitle

\begin{abstract}

In this study we demonstrate a novel Brain Computer Interface (BCI) approach to detect driver distraction events to improve road safety.
We use a commercial wireless headset that generates EEG signals from the brain. 
We collected real EEG signals from participants who undertook a 40-minute driving simulation and were required to perform different tasks while driving. 
These signals are segmented into short windows and labelled using a time series classification (TSC) model.
We studied different TSC approaches and designed a Long-term Recurrent Convolutional Network (LCRN) model for this task.
Our results showed that our LRCN model performs better than the state of the art TSC models at detecting driver distraction events. 

\end{abstract}

\section{Introduction}
Every year, nearly 1.25 million people die from car accidents. On average 3,287 die every day and the numbers are still rising \cite{asirt}. 
According to two news sources, distracted driving is the major cause of car accidents for the past decades \cite{businessinsider,huffpost}. 
Drivers are now more easily distracted than ever and frequently stop paying attention to the road in response to mobile devices, navigation systems and complex control systems \cite{businessinsider}.
It is predicted that road traffic injuries will become the fifth leading cause of death by 2030 if no actions are taken \cite{asirt}.
Therefore, it is important to detect driver distraction events. 

Various video (eye tracking) and speech processing approaches have been proposed to detect driver distraction \cite{vicente2015driver,jo2011vision,mbouna2013visual,yang2011detecting}.
These approaches are sometimes infeasible.
For example,
video processing approaches are less effective with insufficient lighting especially at night when the driver has a higher chance of being less focused on the road.
In car racing environments, drivers are required to wear fire protective suits and helmets that cover their faces, making video processing methods ineffective \cite{salehicar}.
High levels of background noise during driving, for example sound from the engine, radio or wind noise, may decrease the effectiveness of sound processing approaches.

\begin{figure}
    \centering
    \includegraphics[width=\columnwidth]{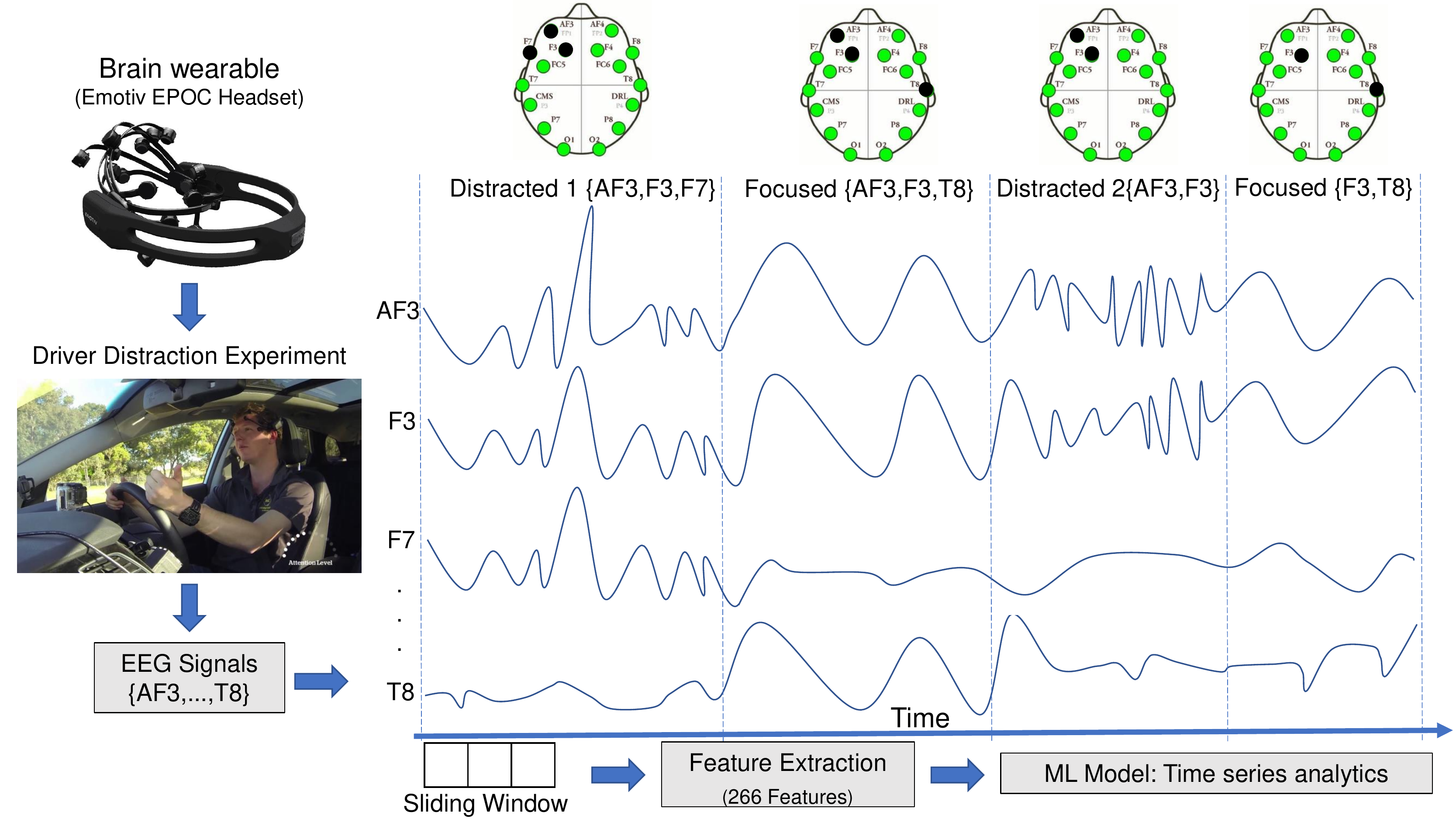}
    \caption{Overview of our driver's distraction detection framework}
    \label{fig:system overview}
\end{figure}

Hence, in this work, we designed a framework to detect driver distraction events with a Brain Computer Interface (BCI) approach, illustrated in Figure \ref{fig:system overview}.
We use a commercial Electroencephalogram (EEG) headset from Emotiv, that is wireless and can easily be worn by a driver.
The EEG data from the device is split into 11.75s segments of time series data using a sliding window and classified using multivariate time series classification (TSC) techniques into one of the two states, the driver is \emph{focused} or \emph{distracted}. 
To the best of our knowledge, none of the previous approaches detect episodes of driver distraction using time series analytics on EEG data \cite{kaplan2015driver}.

\section{Driver distraction detection} 
Our driver distraction detection framework\footnote{\url{https://github.com/ChangWeiTan/EmotivDriverDistraction}} consists of two stages, illustrated in Figure \ref{fig:system overview}. 
First we extract 266 features from the EEG signal.
Then we segment the EEG signal features into short subsequences and train our LCRN model to classify each of the segments.

\subsection{Stage 1: Feature extraction}
We use the lightweight and wireless Emotiv EPOC headset\footnote{http://www.emotiv.com/epoc.php} to monitor the driver's brain in real-time with 14 EEG channel.
In this work, we recorded the EEG signal data from a 40 minutes driving simulation completed by 18 participants. 
Each of them was asked to do 16 tasks (which are the labels of the EEG signal) such as using a phone for text or voice calls, talking to a passenger and solving maths questions to simulate distracted driving. 
Note that 12 participants were randomly selected to be in the training set, 2 in validation set and the remaining were used to evaluate our models.

The raw 14 channel EEG signals generated at 128Hz were processed before passing into a time series classifier. 
A 4-40Hz band-pass filter was applied to reduce ocular artifacts.
Then, Fast Fourier Transform (FFT) was applied to every 2 seconds window with 0.25 second intervals. 
Each of the channels was further split into five frequency bands: \emph{theta, alpha, low beta, high beta and gamma}. 
Then features such as the average power, peak power and peak frequency are extracted from each of the frequency bands. 
Additional features were also generated based on the accumulated power in the left and right frontal regions, left and right hemispheres, left and right temporal/parietal region and occipital power within each specified band, and by accumulating EEP power over the main processing bands (beta and gamma). 
This results in 266 features per EEG signal.

\subsection{Stage 2: Multivariate Time Series Classification}
The state of a driver at any given point in time is detected using multivariate time series classification (TSC), where the objective is to classify every segment of the EEG signal as \emph{focused} or \emph{driving}. 
The 266 dimensional EEG feature set, updated at 4Hz, was segmented into short subsequences of 11.75s windows with 50\% overlaps, creating subsequences with a length of 40 samples.
Each subsequence was labelled with the majority label in that segment and then translated into the state of the driver, i.e., \emph{focused} and \emph{driving}.
A multivariate TSC model is trained to classify each of these subsequences.
In this work, we investigated the feasibility of the following multivariate TSC models for this task:
\begin{enumerate}
    \item \textbf{Euclidean1NN}: One nearest neighbour classifier with Euclidean distance. This is one of the baseline for multivariate TSC \cite{bagnall2018uea}. 
    \item \textbf{Rocket}: The fastest and most accurate univariate TSC algorithm to date \cite{dempster2019rocket}. The original univariate algorithm was adapted for multivariate case by the authors.
    \item \textbf{Residual Network (ResNet)}: ResNet is the most accurate Deep Learning model for univariate TSC \cite{fawaz2019deep}
    \item \textbf{Fully Convolutional Network (FCN)}: FCN is the most accurate Deep Learning model for multivariate TSC \cite{fawaz2019deep}
\end{enumerate}

These models were evaluated using classification accuracy, and $F_1$ scores for both distracted and driving state. 
For the safety of the driver, it is more important to predict the distracted state with high accuracy compared to predicting driving state. 
However, our initial results, illustrated in Figure \ref{fig:results} showed that these models were not able to predict the distracted state very well.
The best overall classifier was Rocket with an accuracy of 0.63 and $F_1$ score of 0.45 and 0.72 for distracted and driving state respectively.
Although ResNet had the highest accuracy, it had the worst distracted $F_1$ score of 0.22.
Hence, we designed a LRCN model to improve our framework \cite{donahue2015long}.

\subsubsection{Our LRCN Model}
\begin{figure}
    \centering
    \includegraphics[width=\columnwidth]{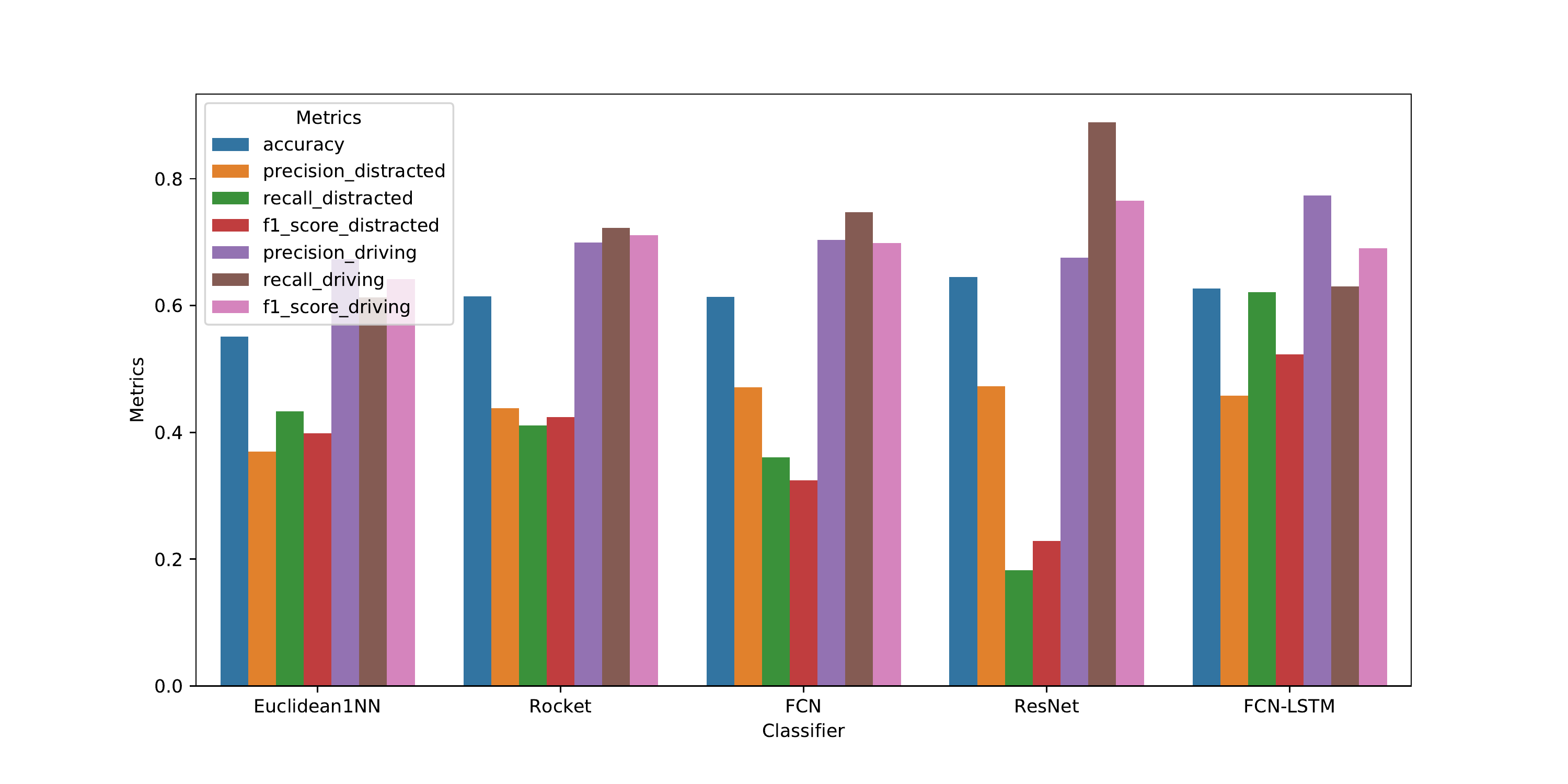}
    
    \caption{Results comparing the different multivariate TSC models averaged over 5 repetitions}
    \vspace*{-10pt}
    \label{fig:results}
\end{figure}

The intuition behind our proposed model is to extract features from the subsequences using the convolutional layers and to learn the relationships between current and past subsequences using a LSTM layer.
LCRN has achieved good performance in other human activity recognition research \cite{sun2018sequential,donahue2015long}.
Our LCRN model utilises the FCN layers as the convolutional layer because it performs better than ResNet in the previous experiment (Figure \ref{fig:results}).
The FCN layers consist of 3 convolutional layers with 128, 256 and 128 filters respectively \cite{fawaz2019deep,wang2017time}.
It is then followed by two LSTM layers and a fully connected softmax layer at the end.
We trained our LRCN (we call it FCN-LSTM) model with 4 subsequences of length 40, each passed into a FCN for feature extraction.

As shown in Figure \ref{fig:results}, our FCN-LSTM network was clearly able to predict the distracted state significantly better than the other models in terms of precision, recall and $F_1$ scores.
We observe that the $F_1$ score for distracted state can still be improved and we believe that this is due to imbalanced class in the dataset and the lack of data.
The current dataset was only generated by 18 participants with only 12 of them being in the training set.
The total training set consists of less than 7000 time series subsequences with only 36\% being in the distracted class.
Furthermore, these subsequences are ``weakly labelled'', meaning that there are extraneous/irrelevant sections in the subsequence that may exist in any other subsequence of different labels \cite{hu2013time}, thus making it extremely challenging to differentiate.

\vspace*{-7pt}
\section{Conclusion}
Overall, we implemented a proof of concept framework to detect driver distraction by classifying EEG signals using a LRCN model.
We found that our LRCN model performs significantly better than the various existing multivariate TSC algorithms for this problem.
However, the performance can still be improved because of the imbalanced class issue and lack of data to generalise well.
In the future, we plan to improve the framework using AutoEncoders as well as adding some post-processing techniques to the framework. 

\subsubsection{Acknowledgements}
We would like to thank Emotiv for providing the Epoc headset and data for this research and Professor Geoff Webb for his comments and guidance.
We are also grateful to the authors of Rocket, FCN and ResNet for providing their code. 

\balance
\bibliographystyle{ieeetr}
\bibliography{references}

\begin{thebibliography}{10}
\bibitem{asirt}
``Top 15 causes of car accidents and how you can prevent them.''
\bibitem{businessinsider}
``Find out what are the leading causes for most car accidents | markets
  insider.''
\end{thebibliography}

%
%
%
%

\end{document}